# Energy Barrier for an Ion Crossing an Intra-Membrane Channel

Alexei V. Finkelstein[1*], Dmitry N. Ivankov[1], Alexander M. Dykhne[2]

[1]Institute of Protein Research, Russian Academy of Sciences, 142290, Pushchino, Moscow Region, Russia;

[2]Center for Theoretical Physics and Applied Mathematics, SSC RF TRINITR, 142190, Troitsk, Moscow Region, Russia

We present a simple approximate analytical estimate for self-energy of a charge in the middle of cylindrical channel of a high permittivity $\varepsilon_1$ in a media of a low permittivity $\varepsilon_2$ (for the cases of infinitely long and comparatively short channels) and show that this estimate is in a good quantitative agreement with exact solution of Poisson equation. Further, using these estimates, we explain the observed a lower conductivity, caused by an increased the self-free-energy for ions, whose diameter is by ~1Å less than that of the channel (as compared to ions, whose diameter is equal to that of the channel).

A pure lipid membrane is virtually impermeable for charges coming from water (having high permittivity $\varepsilon_1 \sim 80$) because of a low permittivity ($\varepsilon_2 \sim 2$) of membrane's inner hydrocarbon part, which is ~ 50 Å thick [1]. Low permittivity of the hydrocarbon layer leads to a very high (by hundreds kJ/mole) increase in electrostatic energy of an ion in a hydrocarbon environment [2]. Therefore, ions cross the membrane via water-filled channels formed by surrounding proteins [1].

A problem of the charge energy inside a channel has been addressed, and a formula for its potential, using an integral of Bessel function, has been obtained [3]. At this basis, the energy of a charge in an infinitely long channel has been calculated as a function of $\varepsilon_2/\varepsilon_1$ and presented as a plot [2]. Later, a numerical solution has been obtained for a channel of finite length [4]. However, obtained solutions have rather complicated form, and, to our best knowledge, no simple (though approximate) equation to estimate the energy barrier experienced by a charge into the middle of the membrane channel has been suggested so far (for the exception of a simplified estimate that one of us has derived and published in a textbook [5] without a mathematical proof).

---

[*] Corresponding author. E-mail: afinkel@vega.protres.ru

Here we would like to derive an approximate expression for the energy of a charge into a channel, and to compare it with precise numerical solution of Poisson equation.

First, let us consider charge $q$ in an infinitely long cylindrical channel of high permittivity $\varepsilon_1$, surrounded by a media of low permittivity $\varepsilon_2$. It is assumed that the channel has radius $a$, that the charge is positioned on the axis of the channel and has radius $b$, that $a$ is sufficiently greater than $b$ (so that water can penetrate between the ion and the channel's wall), and that $\varepsilon_1 \gg \varepsilon_2$. Our aim is to calculate potential $\varphi$ at the surface of the charge and find out the additional energy, which the charge acquires in the channel.

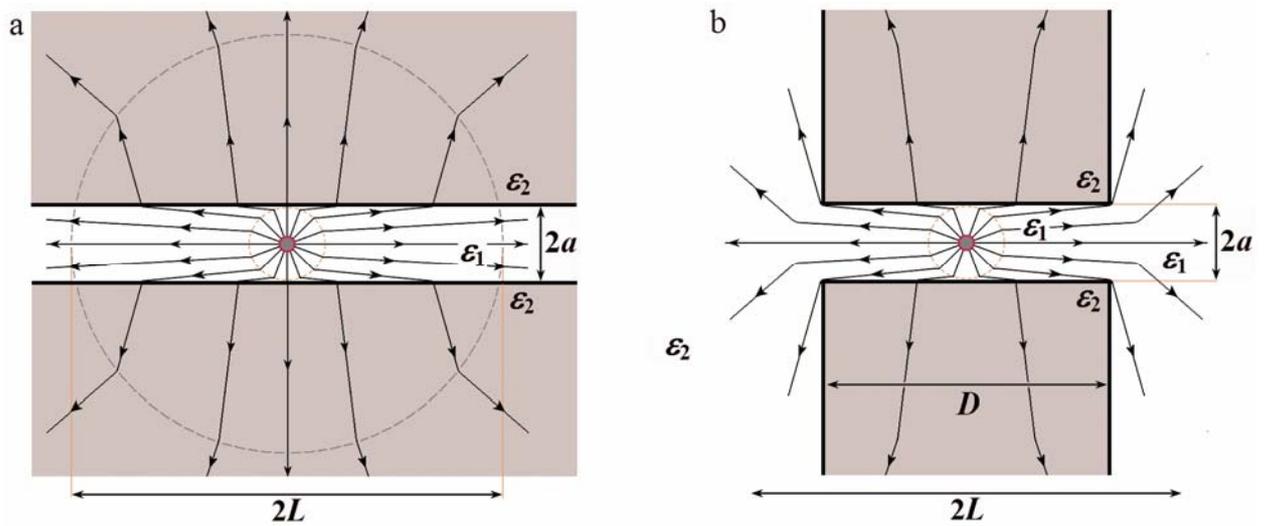

FIG. 1: Electric field of a charge in an infinitely long (a) and short (b) cylindrical channel of high permittivity $\varepsilon_1$, surrounded by a media of low permittivity $\varepsilon_2$; $a$ is radius of the channel, $2L = 2L(\varepsilon_1/\varepsilon_2, a)$ is a critical channel length that separates these two regimes, $D$ is the membrane thickness.

A picture of the field expansion from the charge can be outlined (Fig. 1) using a well known analogy between a flux of force lines of electrostatic field and propagation of electric current: a high-permittivity media is an analog of conductor, and low-permittivity media is an analog of insulator.

Around the charge, electrostatic force lines go at first approximately along radii of the sphere, from the charge surface up to the channel wall; this occurs at such distances $r$, that $b > r > a$.

Far from the charge, at distances $r > L$ from it, these lines again mostly go approximately spherically. The critical distance $L$ will be defined later on; but, evidently, $L \gg a$ when $\varepsilon_1 \gg \varepsilon_2$.

At the intermediate distances $r$ ($a > r > L$), the force lines go mostly along the channel,

because it has high permittivity, but gradually penetrate in the low-permittivity media (Fig. 1a), where the field expands almost cylindrically up to the distance $L$.

The rise of potential at the distances $b > r > a$ is

$$\delta\varphi_{b-a} = q/\varepsilon_1 b - q/\varepsilon_1 a \qquad (1)$$

The rise of potential at the distances $L > r > \infty$ is

$$\delta\varphi_{L-\infty}(L) = q/\varepsilon_2 L = \{q/[a(\varepsilon_1\varepsilon_2)^{1/2}]\}/Z = \Phi/Z \,; \qquad (2)$$

here and below we use $\Phi = q/[a(\varepsilon_1\varepsilon_2)^{1/2}]$ and $Z = (L/a)/(\varepsilon_1/\varepsilon_2)^{1/2}$ (i.e., $L = a(\varepsilon_1/\varepsilon_2)^{1/2} Z$).

The rise of potential at the distances $L > r > a$ is computed as follows.

Along the channel, the electrostatic field intensity is close to $2q/\varepsilon_1 a^2$ at distances $r \sim a$ (because here almost all force lines are in the channel, half of them going one direction through the channel's cross-section of $\pi a^2$, and another half going the opposite direction through the cross-section of same size, see Fig.1); and the field intensity is close to $q/\varepsilon_2 L^2$ at distances $r \sim L$, where the field expansion becomes spherical again [$q/\varepsilon_2 L^2$ must be smaller than $2q/\varepsilon_1 a^2$, of course; this means that $(L/a)^2 > \frac{1}{2}(\varepsilon_1/\varepsilon_2)$, or $Z > 0.5^{1/2}$]. Thus, the average field intensity in the channel's region $a > r > L$ is $\sim (2q/\varepsilon_1 a^2 + q/\varepsilon_2 L^2)/2$, and the total rise of potential along the channel in this region can be estimated as

$$\delta\varphi_{a-L,\text{ in channel}} = [(2q/\varepsilon_1 a^2 + q/\varepsilon_2 L^2)/2](L - a) = (\Phi/Z)(Z^2 + \frac{1}{2})(1 - 1/[Z\times(\varepsilon_1/\varepsilon_2)^{1/2}]) \,.(3)$$

In the low-permittivity media, the perpendicular to the channel electrostatic field intensity is also about $q/\varepsilon_2 L^2$ at distances $r \sim L$, where the field expansion becomes spherical. Since at smaller distances $r$ from the channel's axis the field expands cylindrically, its intensity is about $q/(\varepsilon_2 L r)$, and the total rise of potential in the region of cylindrical expansion ($L > r > a$) is

$$\delta\varphi_{a-L,\text{ perp. to channel}} = (q/\varepsilon_2 L)\ln(L/a) = (\Phi/Z) \ln[Z\times(\varepsilon_1/\varepsilon_2)^{1/2}] \,. \qquad (4)$$

Since $\delta\varphi_{a-L,\text{ in channel}} = \delta\varphi_{a-L,\text{ perp. to channel}}$, one can estimate the $L = a(\varepsilon_1/\varepsilon_2)^{1/2} Z$ value from equation

$$(Z^2 + \frac{1}{2})(1 - 1/[Z\times(\varepsilon_1/\varepsilon_2)^{1/2}]) = \ln Z + \frac{1}{2}\ln(\varepsilon_1/\varepsilon_2) \,. \qquad (5)$$

This equation has two solutions:

$Z = (\varepsilon_1/\varepsilon_2)^{-1/2}$ at all $\varepsilon_1/\varepsilon_2$ values

(6)

$Z \approx [\frac{1}{2}\ln(\varepsilon_1/\varepsilon_2)]^{1/2}$ : another solution that exists at $\varepsilon_1/\varepsilon_2 > 2^{1/2}$ only; it leads to a lower potential, and therefore this is the main solution at small $\varepsilon_2$.

These solutions splice at $\varepsilon_1 = 2.35\varepsilon_2$, where $\frac{1}{2}\ln(\varepsilon_1/\varepsilon_2) = (\varepsilon_1/\varepsilon_2)^{-1}$.

The resulting critical distance $L$ is $a$ at when $\varepsilon_1 \geq \varepsilon_2 \gtrsim 0.5\varepsilon_1$, while at $\varepsilon_2 \ll \varepsilon_1$ it is

$$L(\varepsilon_1/\varepsilon_2, a) \approx a[(\varepsilon_1/2\varepsilon_2)\ln(\varepsilon_1/\varepsilon_2)]^{1/2}. \tag{7}$$

Thus, the total potential acting at the charge is

$$\varphi = \delta\varphi_{b-a} + \delta\varphi_{L-\infty} + \delta\varphi_{a-L,\ in\ channel} = \delta\varphi_{b-a} + (\Phi/Z)\{\ln[Z\times(\varepsilon_1/\varepsilon_2)^{1/2}] + 1\} \tag{8}$$

and the additional (as compared to the bulk media with permittivity $\varepsilon_1$) free energy that the charge acquires in the channel is

$$\Delta U = \varphi q/2 - q^2/2\varepsilon_1 b = (q\Phi/2)[\{\ln[Z\times(\varepsilon_1/\varepsilon_2)^{1/2}] + 1\}/Z - (\varepsilon_1/\varepsilon_2)^{-1/2}]. \tag{9}$$

Thus, the result is:

$$\Delta U = \{q^2/2a\}[1/\varepsilon_2 - 1/\varepsilon_1] \text{ at } \varepsilon_1 \leq 2.35\varepsilon_2;$$

$$\Delta U \approx \{q^2/[2a(\varepsilon_1\varepsilon_2)^{1/2}]\}[\{\ln[Z\times(\varepsilon_1/\varepsilon_2)^{1/2}] + 1\}/Z - (\varepsilon_1/\varepsilon_2)^{-1/2}], \text{ where } Z \approx [\tfrac{1}{2}\ln(\varepsilon_1/\varepsilon_2)]^{1/2}, \tag{10}$$
$$\text{at } \varepsilon_1 \geq 2.35\varepsilon_2.$$

Paradoxically, the simplest solution $Z \approx [\tfrac{1}{2}\ln(\varepsilon_1/\varepsilon_2)]^{1/2}$, obtained for a large $\varepsilon_1/\varepsilon_2$ ratio, turns out to be rather precise also for small $\varepsilon_1/\varepsilon_2$ ratio (provided $\varepsilon_1/\varepsilon_2 > 2^{1/2}$), and the $\Delta U$ value obtained with $Z = [\tfrac{1}{2}\ln(\varepsilon_1/\varepsilon_2)]^{1/2}$ is close (within percents) to the $\Delta U$ value obtained with precise solution $Z$ of equation (5), see Table 1. Also, a strict solution for an infinitely long channel at various $\varepsilon_1/\varepsilon_2$ ratios, based either on integrals of Bessel functions (cf. [2, 3]), or on a numerical solution of Poisson equation are in a fairly good concordance with the approximate analytical estimate (10), see Table 1.

**Table 1**

| $\varepsilon_2$ | $(\varepsilon_1/\varepsilon_2)^{1/2}$ (where $\varepsilon_1$=80) | $Z_{\text{precise}}$ from Eq.(5) | $Z_{\text{approx}} = (\varepsilon_1/\varepsilon_2)^{-1/2}$ from Eq.(6) | $L/a$ from $Z_{\text{precise}}$ | $L/a$ from $Z_{\text{approx}}$ | $\Delta U$ in $q^2/[a(\varepsilon_1\varepsilon_2)^{1/2}]$ units | | | |
|---|---|---|---|---|---|---|---|---|---|
| | | | | | | from Eq.(8) with $Z_{\text{precise}}$ | $[\tfrac{1}{2}\ln(\varepsilon_1/\varepsilon_2)]^{1/2}$ from Eq(10) [or Eq.(8) with $Z_{\text{approx}}$] | Exact analytical solution (from numerical integration of the Smythe's formulae [3]) | Our numerical solution of Poisson equation for Fig.1a |
| 80. | 1.000 | 1.00 | 1.00 | 1.0 | 1.0 | 0.000 | 0.000 | 0 | 0 |
| 40. | 1.413 | 0.71 | 0.71 | 1.0 | 1.0 | 0.353 | 0.353 | 0.28 (from plot in [2]) | 0.257 |
| 10. | 2.828 | 1.08 | 1.02 | 3.0 | 2.9 | 0.805 | 0.833 | 0.70 (from plot in [2]) | 0.694 |
| 2. | 6.325 | 1.40 | 1.36 | 8.9 | 8.6 | 1.057 | 1.081 | 1.08 (data from [6]) | 1.069 |
| 1. | 8.944 | 1.52 | 1.48 | 13.6 | 13.2 | 1.130 | 1.154 | 1.20 (from plot in [2]) | 1.185 |

For a water channel in a membrane, estimate (10) should be valid when the ratio of the channel diameter $2a$ to the membrane thickness $D$ is much less than $a/L = [(\varepsilon_1/2\varepsilon_2)\ln(\varepsilon_1/\varepsilon_2)]^{-1/2} \approx 0.12$ at $\varepsilon_1 = 80$ and $\varepsilon_2 \approx 2$ (which is typical for permittivity of a water-filled membrane channel [1]). Thus, at $D \approx 50$ Å, which is typical for a membrane [1], equation (8) is valid when the channel diameter $2a$ is below 6 Å.

For a wider channel, the force lines do not penetrate into the membrane, go mainly along the channel and then through bulk water (Fig.1a), and the resulting (cf. equation 3) estimate of the additional energy (which is due to the field expansion through the narrow channel) is

$$\Delta U_{\text{thick channel}} \approx (q^2/\varepsilon_1 a)(D/2a - 1). \tag{11}$$

One can see that this estimate does not depend on $\varepsilon_2$ value; the only requirements are that $\varepsilon_2 \ll \varepsilon_1$ and $a[(\varepsilon_1/2\varepsilon_2)\ln(\varepsilon_1/\varepsilon_2)]^{1/2} > D$. Numerical solution for a wide channel is given in [5].

In conclusion, we should mention the following. The above estimates (10) and (11) do not depend on the ion radius $b$, provided the ion is surrounded by water. These estimates and all the above explanations should also hold when $b = a$, and the ion touches only either water or membrane. However, if $2b$ is less than $2a$ by an angstrom or so, water molecules cannot penetrate between the ion and the channel wall, and the ion is separated from the wall by a layer of vacuum (Fig.2).

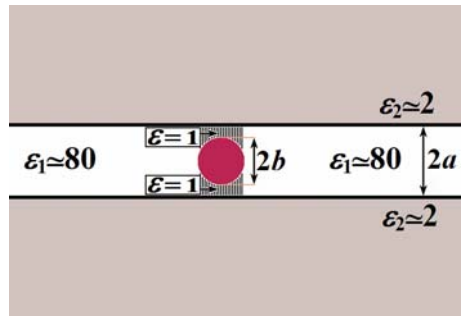

FIG. 2: When $2a - 2b < d_S$ ($d_S$ being diameter of the solvent molecules) the solvent cannot penetrate between the ion and the channel wall, and the ion is separated from the wall by a cylindrical layer of vacuum.

This means that the effective diameter of the channel shrinks from $2a$ to $2b$ at a distance of about $b$, which increases $\Delta U$ by $\approx (q^2/\varepsilon_1 b)(a/b - 1)$ (see equation 11) and explains why the ion permeability, which is sufficiently high for ions whose radius $2b = 2a \approx 3$ Å, but decreases by a couple of orders of magnitude [1] when $2b \approx 2a - 1$ Å.

**Acknowledgements**

This work was supported by the program "Molecular and Cellular Biology" of the Russian Academy of Sciences, by the program "Scientific Schools", by the Russian Foundation for Basic Research, by the company Algodign Ltd and by an International Research Scholar's to A.V.F. from the Howard Hughes Medical Institute.